\documentclass{svproc}
\usepackage{url}
\usepackage{graphicx}
\usepackage{pdflscape}
\usepackage{multicol}
\usepackage{float}
\usepackage{subfig}

\usepackage[bottom]{footmisc}
\begin{document}
\mainmatter              
\title{Strangeness production in p-Pb collisions at \\ 8.16 TeV}
\titlerunning{Strangeness production in p-Pb collisions at 8.16 TeV}  
%
\author{Meenakshi Sharma, for ALICE Collaboration}
\authorrunning{Meenakshi Sharma for the ALICE Collaboration} 
%

%
\institute{Department of Physics, University of Jammu, Jammu and Kashmir 180006, India\\
\email{meenakshi.sharma@cern.ch}\\ 
}

\maketitle              

\begin{abstract}

The analysis status of strange hadrons ($K^{0}_{s}$ and $\Lambda$ ) in p-Pb collisions in multiplicity bins is presented as a function of \textit{p}$\rm_{T}$  for -0.5 $<$ $y_{\rm CMS}$ $<$0. The excellent tracking and particle identification capabilities of ALICE can be used to reconstruct the strange hadrons using the tracks produced by their weak decays. The production rate of strange quarks is one of the various observables sensitive to the evolution of the system after nuclear collisions. It is now confirmed that strange quarks would be produced with higher probability in a QGP scenario with respect to that expected in a pure hadron gas scenario. Therefore, studies of strangeness production can help to determine the properties of the created system.

\end{abstract}

\section{Introduction}

Strangeness production plays a key role in the study of hot and dense systems created in nuclear collision. Rafelski and Muller \cite{ref:FONLL} reported for the first time that the enhancement of the relative strangeness production could be one of the signatures of a phase transition from hadronic matter to the new phase consisting of almost free quarks and gluons (QGP). Strangeness enhancement was observed in several experiments [2-4] as a function of number of participating nuclei $<N_{\rm part}>$. The enhancement is relative to the production of strangeness in small systems (pp or p-Be) where the enhanced production was not expected.
Recent studies by the ALICE collaboration show that the strange particle yield with respect to pion yield increases smoothly with increase in $\left\langle \rm dN_{ch} \right\rangle   /\left\langle d\eta \right\rangle $ [5,6]. 
\section{Detection of strange hadrons in p-Pb collisions with ALICE}
 With an overall branching ratio of 69.2\% (63.9\%),  $\rm K^{0}_{s} $ ($\rm \Lambda $) hadrons decay weakly into $\pi^{+}\pi^{-}$ (p$\pi^{-}$) pairs. The tracks formed by the daughter particles of the decays are reconstructed in Inner Tracking System (ITS) and in the large Time Projection Chamber (TPC). The particle identity is determined through the measured ionisation energy loss and momentum measured in the TPC. The secondary vertex reconstruction from particle decays is performed as shown in Fig. 1.
 \begin{figure}[h!]
\centering
\includegraphics[scale=0.2]{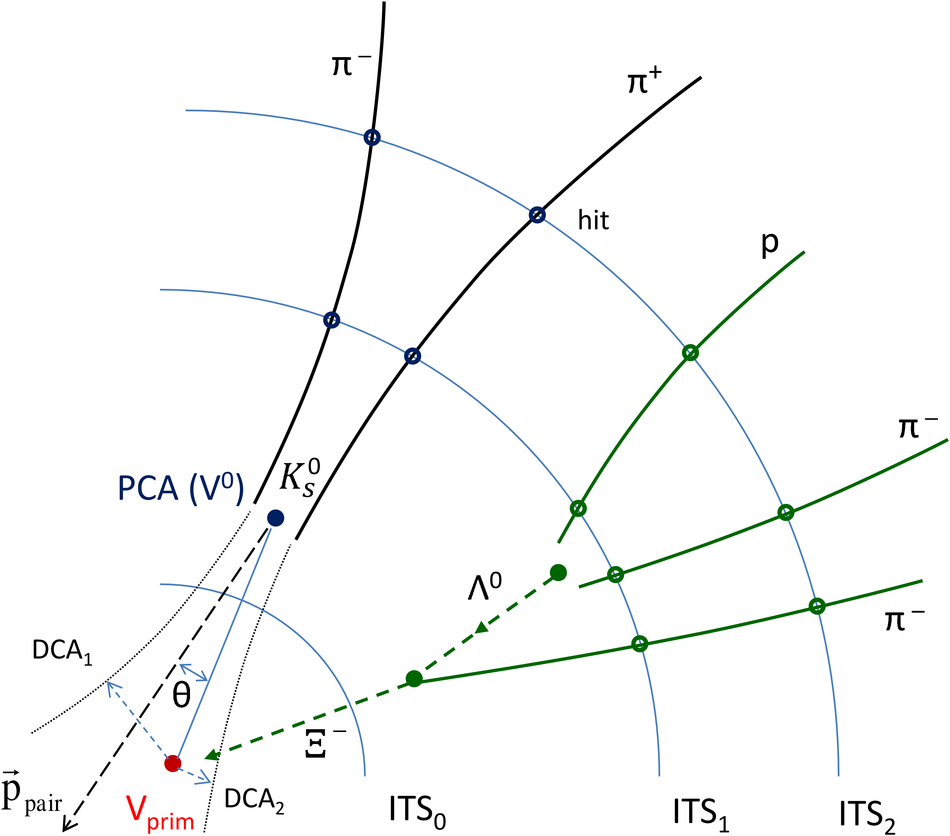} 
\caption{Secondary vertex reconstruction principle, with  $\rm K^{0}_{s} $ and $\rm\Xi^{-}$ decays shown as examples. For clarity, the decay points are placed between the first two ITS layers (radii are not to scale). The solid lines represent the reconstructed charged particle tracks, extrapolated to secondary vertex. Extrapolations to the primary and secondary vertices are shown in dashed lines. }

\end{figure} 

To reduce the combinatorial background in the selection of the weakly decaying particles, a set of geometrical cuts based on the decay topologies were applied. These cuts were imposed on the minimum Distance of Closest Approach (DCA) between the V0 daughter tracks and the primary vertex. Another cut was applied on the DCA between charged daughter tracks.
\section{Analysis}
A detailed description of the ALICE detector can be found in \cite{ref:Percolation}. The data used for the analysis was recorded in 2016. The particle identification is done using following ALICE sub detectors: the Inner Tracking System (ITS), the Time Projection Chamber (TPC), by using different PID techniques \cite{ref:detector}. The multiplicity bins are defined based on the signal amplitude measured in the V0A scintillator. \\
    \par The signal was extracted from the invariant mass distribution of the decay daughters. In every \textit{p}$\rm_{T}$ bin, a Gaussian function was fit to the mass peak and a central area defined $\rm -4\sigma$ and $\rm 4\sigma$ for that peak. The sigma is in the range 0.00395 to 0.00404 for the four ranges of \textit{p}$\rm_{T}$. The candidate counts in two background bands defined in the intervals [-10;-6]$\rm\sigma$ and [6;10]$\rm\sigma$ from the Gaussian peak were subtracted from the entries in the central bins to give the signal.
 
\begin{figure}[h!]

\includegraphics[scale=0.38]{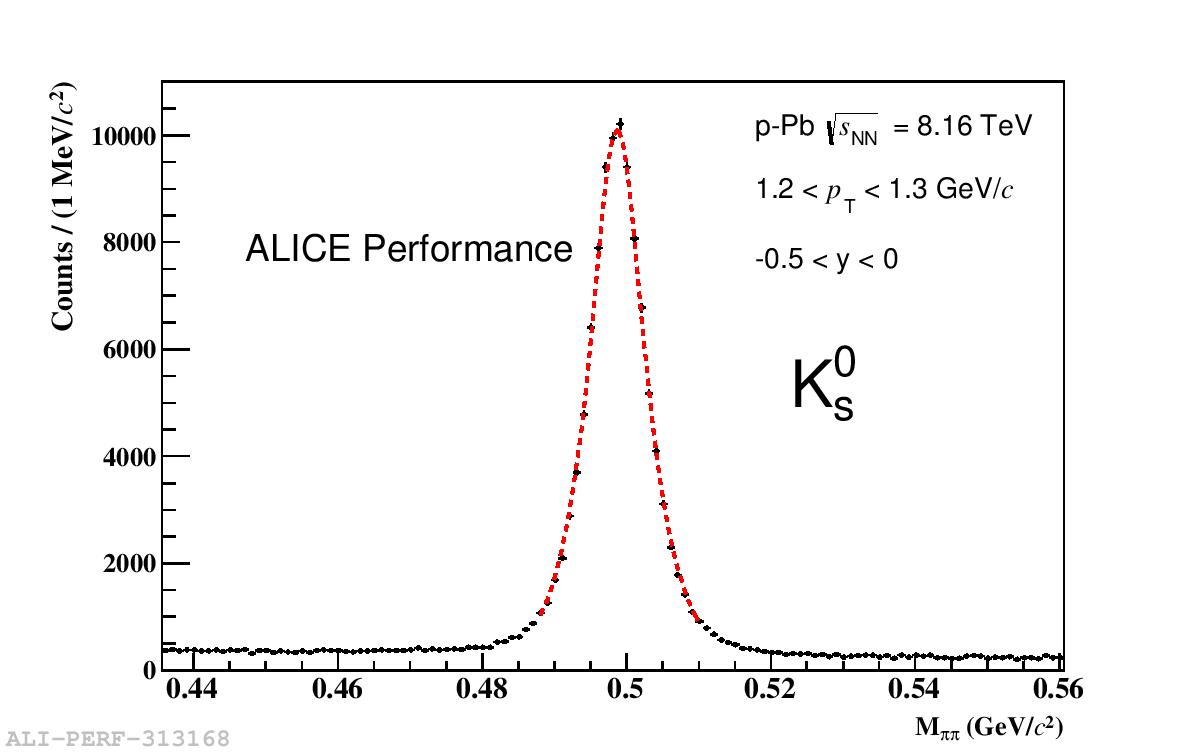} \includegraphics[scale=0.38]{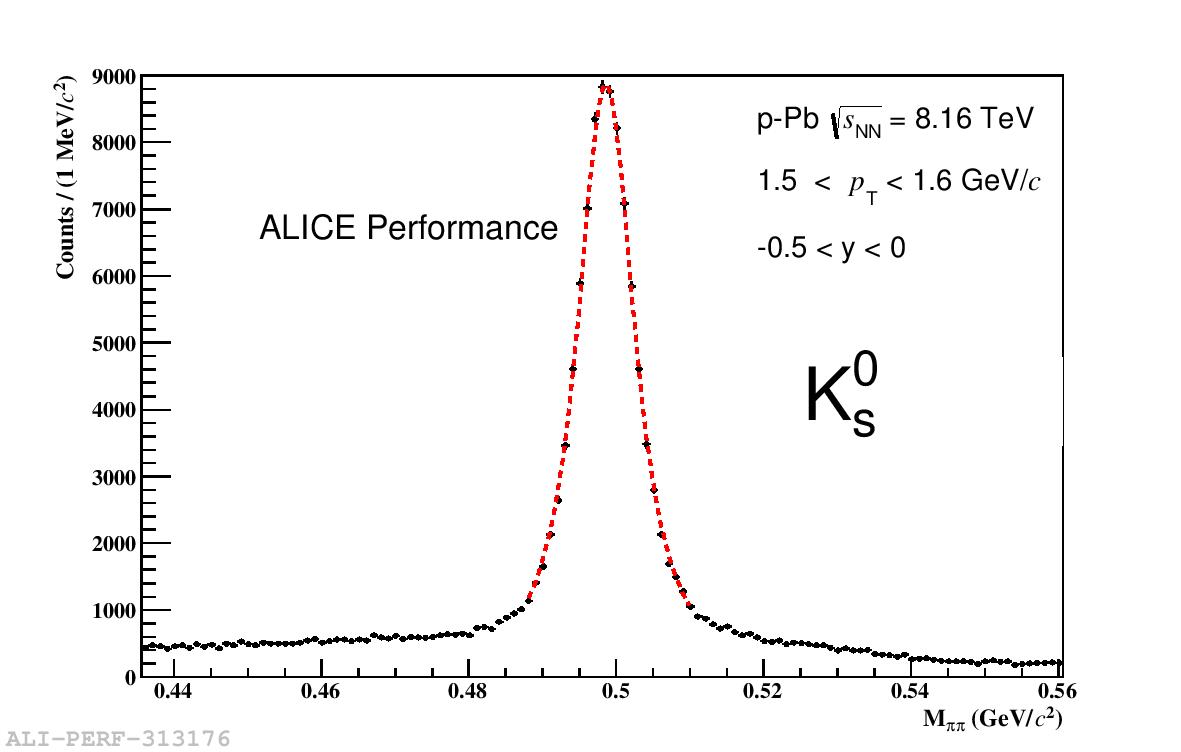}
\includegraphics[scale=0.38]{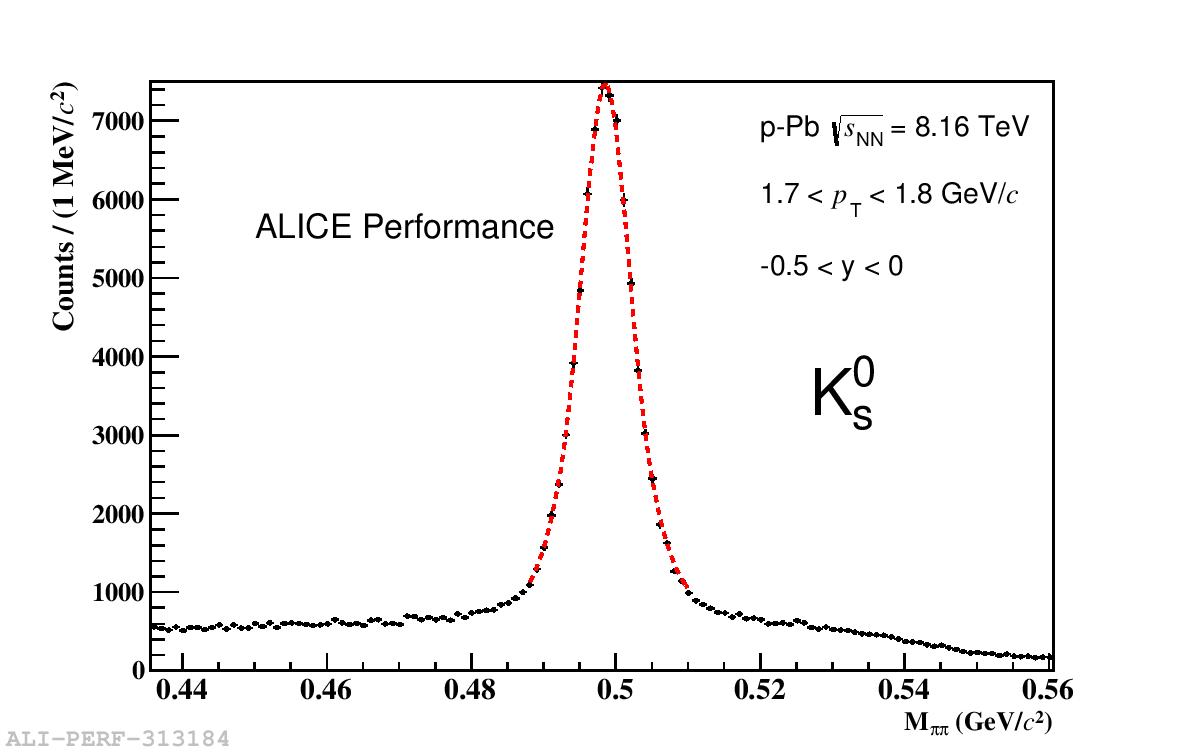} \includegraphics[scale=0.38]{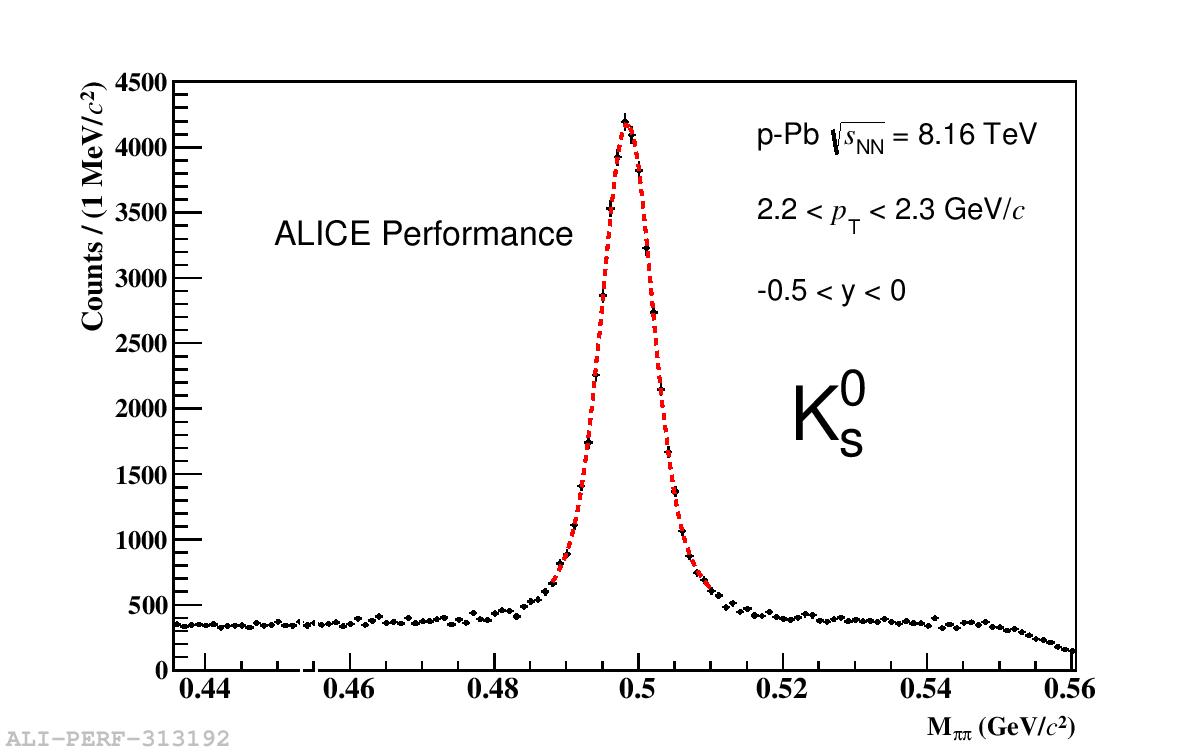}
\caption{Invariant mass distributions of $\rm K^{0}_{s} $ in four \textit{p}$\rm_{T}$ bins 1.2-1.3 GeV/c, 1.5-1.6 GeV/c, 1.7-1.8 GeV/c and 2.2-2.3 GeV/c, fitted with Gaussian peaks(dashed red curves) and linear backgrounds.The distributions are for the 0-100\% multiplicty class.}
\end{figure}


\section{Outlook}

 The invariant mass distributions have been presented in Figure 2 in four different \textit{p}$\rm_{T}$ bins for the 0-100\% multiplicty class. Figure 3 shows the comparison of the yields of various hadrons relative to pions in pp collisions at $\sqrt{s}$ =13 TeV, Pb-Pb collisions at $\sqrt{s_{\rm NN}}$ = 5.02 TeV, pp collisions at $\sqrt{s}$ = 7 TeV, p-Pb collisions at $\sqrt{s_{\rm NN}}$ = 5.02 TeV and Xe-Xe collisions at $\sqrt{s_{\rm NN}}$ = 5.44 TeV.\\

 \begin{figure}[h!]
\centering

\includegraphics[scale=0.11]{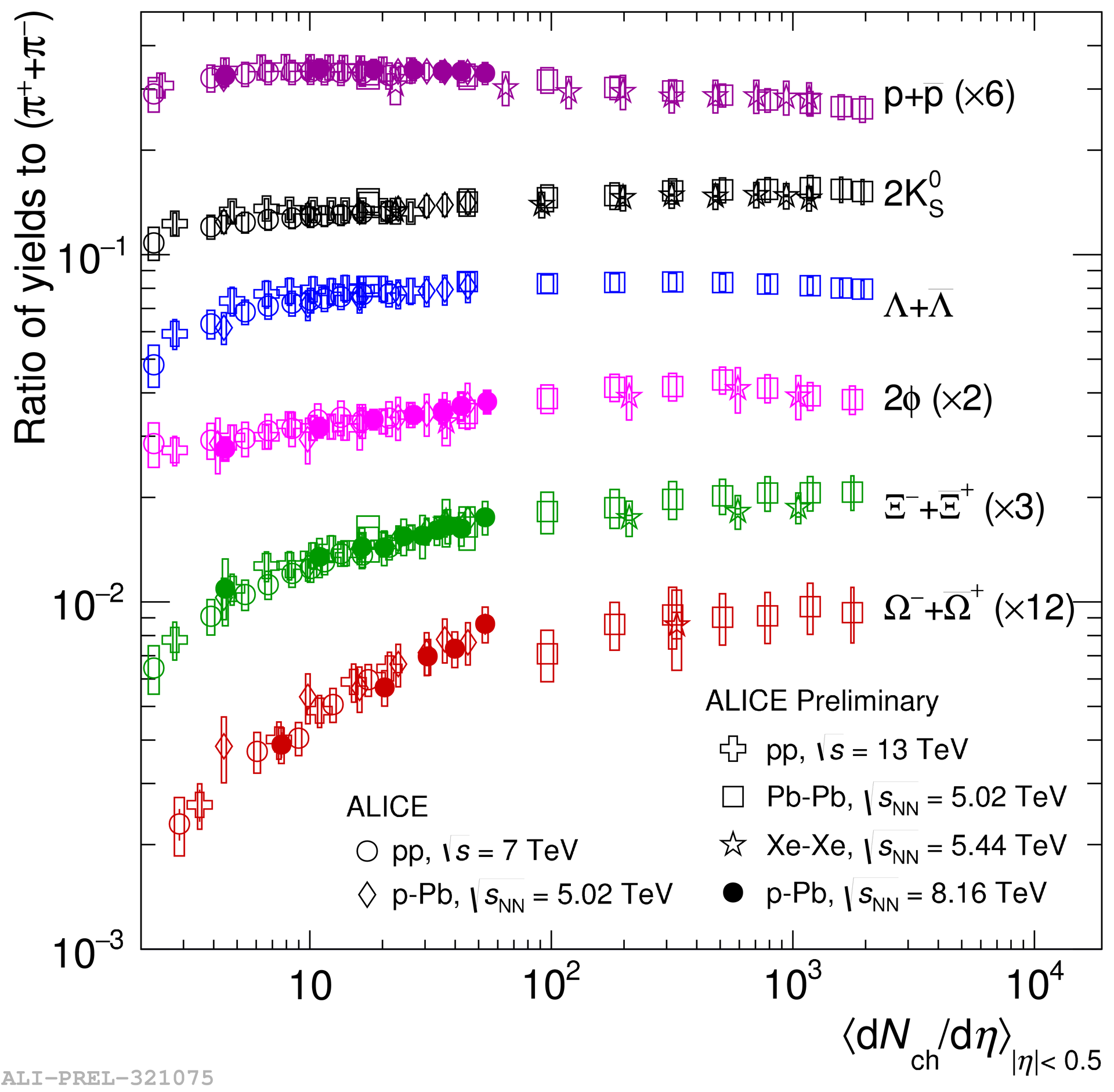} 
\caption{Comparison of yields of protons, $\rm K^{0}_{s} $, ($\rm \Lambda + \overline \Lambda)$, $\rm \phi$, $\rm (\Xi^{-} + \Xi^{+})$ and $\rm (\Omega^{-} + \overline{\Omega} ^{+})$  relative to pion yields in pp collisions at $\sqrt{s}$ =13 TeV, Pb-Pb collisions at $\sqrt{s_{\rm NN}}$ = 5.02 TeV, pp collisions at $\sqrt{s}$ = 7 TeV, p-Pb collisions at $\sqrt{s_{\rm NN}}$ = 5.02 TeV, Xe-Xe collisions at $\sqrt{s_{\rm NN}}$ = 5.44 TeV and p-Pb collisions at $\sqrt{s_{\rm NN}}$ = 8.16 TeV }
\end{figure} 
 
\par From this comparison one can see that the points from different systems but with similar value of $\left\langle \rm dN_{ch}  /d\eta \right\rangle$ overlap. This suggests that relative strangeness production does not depend on the initial stage parameters such as identity of the colliding ions or collision energy, but depends on the final state charged particle density. The $\left\langle \rm dN_{ch}  /d\eta \right\rangle$ covers three orders of magnitude starting with strong strangeness enhancement for multistrange particles at low value of $\left\langle \rm dN_{ch}  /d\eta \right\rangle$ (possible canonical suppression scenario) with constant relative production of strangeness for high values of $\left\langle \rm dN_{ch}  /d\eta \right\rangle$ (possible grand canonical saturation scenario). The present analysis is a preliminary work and measurements of $\rm K^{0}_{s} $ and $\rm \Lambda $ in p-Pb collisions at  $\sqrt{s_{\rm NN}}$ = 8.16 TeV will be added to Fig. 3. This will allow us to see whether p-Pb system at 8.16 TeV shows the same trend as the other mentioned systems or not.
\\
\vspace{-0.80cm}
%

\end{document}